\documentstyle[12pt]{article}

\begin{document}

\author{C. S. Unnikrishnan\thanks{%
\noindent E-mail: unni@tifr.res.in} \\
{\it Gravitation Group, Tata Institute of Fundamental Research, }\\
{\it \ Homi Bhabha Road, Mumbai - 400 005 }\\
{\it \ and }\\
{\it \ NAPP Group, Indian Institute of Astrophysics, }\\
{\it \ Koramangala, Bangalore - 560~034}}
\title{Three-particle GHZ correlations without nonlocality}
\date{}
\maketitle

\begin{abstract}
The formalism employing local complex amplitudes that resolved the
Einstein-Podolsky-Rosen puzzle (C. S. Unnikrishnan, quant-ph/0001112) is
applied to the three-particle GHZ correlations. We show that the GHZ quantum
correlations can be reproduced without nonlocality.
\end{abstract}

We have recently shown that quantum correlations can be reproduced starting
from local probability amplitudes, by calculating the correlations from
amplitudes directly rather than by multiplying the outcomes and integrating
over some hidden variable values \cite{unni1}. In the local hidden variable
theories the correlations are calculated from eigenvalues and this procedure 
{\em does not preserve the phase information}. The situation has some
analogy to the description of interference in quantum mechanics. Any attempt
to reproduce the interference pattern using locality and the information on
`which-path' will fail since the phase information is lost or modified in
such an attempt. In the local amplitude formalism, measurement on one
particle does not cause the companion particle to acquire, or to collapse
into, a definite state. (The present interpretation of quantum teleportation
and entanglement swapping will change in this local picture, without
affecting the actually measured correlations).

The statement of locality is at the level of the probability amplitudes and
can be written as 
\begin{equation}
C_{1\pm }=C_{1\pm }({\bf a},\phi _{1}),\quad C_{2\pm }=C_{2\pm }({\bf b}%
,\phi _{2})
\end{equation}
where ${\bf a}$ and ${\bf b}$ are local settings for analyzers, and $\phi
_{1}$ and $\phi _{2}$ are the internal variables (`hidden variables')
associated with the individual particles and appear in the amplitudes as a
phase \cite{unni1}. A definite value for these variables does not imply a
define state for the particles before the measurement.

The locality assumption also implies the locality for observables $A$ and $B$%
,

\begin{equation}
A({\bf a,}\phi _{1})=\pm 1,\quad B({\bf b,}\phi _{2})=\pm 1
\end{equation}

This is the same locality assumption as in local realistic theories. But,
this has a meaning different from its meaning in standard local realistic
theories. Here, this means that the outcomes, {\em when measured}, depend
only on the local setting and the local internal variable. There is no
objective reality to $A$ and $B$ before a measurement. There is objective
reality to $\phi _{1}$ and $\phi _{2},$ but there is no way to observe these
absolute phases.

In this framework the correlation function is not $P({\bf a},{\bf b})=\frac{1%
}{N}\sum (A_{i}B_{i})$ or $\int d{\bf \phi }\rho ({\bf \phi })A({\bf a,}\phi
_{1})B({\bf b,}\phi _{2}).$ The correct correlations are of the form, 
\begin{equation}
U({\bf a},{\bf b})=~Real(NC_{i}C_{j}^{*})
\end{equation}
where $N$ is a normalization factor. It is the square of this correlation
function that would give a joint probability. The correlation of the
eigenvalues $P({\bf a},{\bf b})=\frac{1}{N}\sum (A_{i}B_{i})$ also can be
derived from the absolute square of $U({\bf a},{\bf b})$ \cite{unni1}. {\em %
The crucial difference from local realistic theories is that the correlation
is calculated from quantities which preserve the relative phases. }

We now apply this formalism for the description of correlations of the three
particle G-H-Z state \cite{ghz} defined as 
\begin{equation}
\left| \Psi _{GHZ}\right\rangle =\frac{1}{\sqrt{2}}(\left|
1,1,1\right\rangle -\left| -1,-1,-1\right\rangle )
\end{equation}

where the eigenvalues in the kets are with respect to the $z$-axis basis.

The conflict between a local realistic theory and quantum mechanics is the
following statement \cite{ghz}:

The prediction from quantum mechanics for the measurement represented by the
operator $\sigma _{x}^{1}\otimes \sigma _{x}^{2}\otimes \sigma _{x}^{3}$ is
given by 
\begin{equation}
\sigma _{x}^{1}\otimes \sigma _{x}^{2}\otimes \sigma _{x}^{3}\left| \Psi
_{GHZ}\right\rangle =-\left| \Psi _{GHZ}\right\rangle
\end{equation}

Local realistic theories predict that the product of the outcomes in the $x$
direction for the three particles should be $+1.$ This contradicts Eq. 5.

We now show that the quantum prediction can be reproduced using local
amplitudes. The general idea is that the three particle correlation,
analogous to our scheme for two-particle states \cite{unni1}, is the real
part of a complex number $Z$ obtained as a suitable product of three complex
amplitudes. We choose the different phases such that the correlation
represented by Real$(Z---)$ is $\pm 1$ (i.e. $(Z---)$ is pure real) to
satisfy the condition that the joint probability for the outcome $(-,-,-)$
is unity according to Eq. 5. The rest of the correlations follow without any
additional input since flipping the sign once (for example Real($Z+--)$)
amounts to rotating $Z$ through the phase $\pi /2.$ This is because the
amplitudes for $+$ and $-$ are orthogonal. The joint probability itself is
the square of the correlation function and clearly these joint probabilities
are unity for the outcomes containing an odd number of $(-).$ 

We define the local amplitudes for the outcomes $+$ and $-$ at the analyzer
(with respect to the $x$ basis) for the first particle as $C_{1+}=\frac{1}{%
\sqrt{2}}\exp (i\theta _{1})$, and $C_{1-}=\frac{1}{\sqrt{2}}\exp (i(\theta
_{1}+\pi /2)).$ The amplitude $C_{1-}$ contains the added angle $\pi /2$
because this amplitude is orthogonal to $C_{1+}.$ Similarly, we have $C_{2+}=%
\frac{1}{\sqrt{2}}\exp (i\theta _{2})$, and $C_{2-}=\frac{1}{\sqrt{2}}\exp
(i(\theta _{2}+\pi /2))$ for the second particle and $C_{3+}=\frac{1}{\sqrt{2%
}}\exp (i\theta _{3})$, and $C_{3-}=\frac{1}{\sqrt{2}}\exp (i(\theta
_{3}+\pi /2))$ for the third particle. Our aim is to choose the various
phases such that the following is true: 
\begin{eqnarray}
P(+,+,+) &=&0  \nonumber \\
P(-,-,-) &=&1  \nonumber \\
P(+,+,-) &=&1  \nonumber \\
P(+,-,+) &=&1  \nonumber \\
P(-,+,+) &=&1  \nonumber \\
P(-,-,+) &=&0  \nonumber \\
P(+,-,-) &=&0  \nonumber \\
P(-,+,-) &=&0
\end{eqnarray}

These are the quantum mechanical predictions for the joint probabilities for
getting the outcomes indicated.

We choose the following definition for the correlation function whose square
is the relevant joint probability. (The final results are independent of the
particular definition we use. Once a definition is chosen the phases can be
solved for the outcomes).

Correlation function is obtained from the definition $N$Real($%
C_{1}C_{2}^{*}C_{3}^{*}),$ where $N$ is a normalization constant. Since we
want $N$Real($C_{1-}C_{2-}^{*}C_{3-}^{*})=\pm 1,$ we choose $%
C_{1-}C_{2-}^{*}C_{3-}^{*}$ to be pure real$.$ This gives 
\[
\frac{N}{2\sqrt{2}}Real(\exp i(\theta _{1}-\theta _{2}-\theta _{3}-\pi
/2))=\pm 1 
\]
\[
\theta _{1}-\theta _{2}-\theta _{3}-\pi /2=0 {\rm {\ or }\pm \pi }
\]
We can choose the relevant relative phases to satisfy this condition. Then
we get 
\[
P(-,-,-)=1 
\]

Rest of the joint probabilities given in Eq. 6 automatically follow, since
flipping sign once rotates the complex number $C_{1-}C_{2-}^{*}C_{3-}^{*}$
through $\pi /2.$ The square of $N$Real($C_{1}C_{2}^{*}C_{3}^{*})$ is then $%
1 $ for an odd number of $(-)$ outcomes and $0$ for even number of $(-)$
outcomes.

This completes the construction of local amplitudes for the three particle
maximally entangled state. Similar construction also applies to four-
particle maximally entangled state \cite{sam} and general multiparticle
maximally entangled states.

\end{document}